\documentclass[aps,prd,preprint,nofootinbib]{revtex4}
\usepackage{bbm}
\usepackage{amsfonts}
\usepackage{mathrsfs}
\usepackage{graphicx}
\usepackage{amsmath}
\usepackage{amssymb}
\usepackage{bm}
\usepackage{bbm}
\usepackage{multirow}
\usepackage{epsfig}
\usepackage{graphicx}
\newcommand{\bqa}{\begin{eqnarray}}
\newcommand{\eqa}{\end{eqnarray}}
\newcommand{\beq}{\begin{equation}}
\newcommand{\eeq}{\end{equation}}
\newcommand{ \slashchar }[1]{\setbox0=\hbox{$#1$}   
   \dimen0=\wd0                                     
   \setbox1=\hbox{/} \dimen1=\wd1                   
   \ifdim\dimen0>\dimen1                            
      \rlap{\hbox to \dimen0{\hfil/\hfil}}          
      #1                                            
   \else                                            
      \rlap{\hbox to \dimen1{\hfil$#1$\hfil}}       
      /                                             
   \fi}                                             %
\begin{document}
\title{Reconstructing the 125GeV SM Higgs boson
through $\ell\bar{\ell}\gamma$\\[7mm]}
\author{Long-Bin Chen$^{1}$\footnote{E-mail:
chenglogbin10@mails.gucas.ac.cn}, Cong-Feng
Qiao$^{1,2}$\footnote{E-mail: qiaocf@gucas.ac.cn }, and Rui-Lin
Zhu$^{1}$\footnote{E-mail: zhuruilin09@mails.gucas.ac.cn}}


\address{$^{1}$Department of Physics,
Graduate University of the Chinese Academy of Sciences,
\\YuQuan Road 19A, Beijing 100049, China\\
$^{2}$Kavli Institute for Theoretical Physics China, the Chinese
Academy of Sciences, Beijing 100190, China}
\author{~\vspace{0.7cm}}

\begin{abstract}

\vspace{3mm}
To ascertain the new boson with mass near 125 {GeV}
observed recently by ATLAS and CMS Collaborations to be the
Standard Model Higgs, and to determine its intrinsic properties,
more measurements on its various decay channels
are still necessary. In this work we reanalyze the
processes of the Standard Model Higgs radiative decays to
lepton pairs. We find that when photon and leptons are hard, that is
possessing energies larger than 1 {GeV}, the branching fractions of
$H\rightarrow \ell\bar{\ell}\gamma (\ell=e~or~\mu)$ processes are
about two-thirds of the $H\rightarrow\mu^+\mu^-$ process. Since the
lepton-pair yields of the radiative processes mainly come from the
Z-boson conversion, which will greatly suppress the backgrounds, we
believe the signal should be observable in presently accumulated
data or in the next run of the LHC experiment, provided the Standard
Model Higgs is indeed light.

\vspace {7mm} \noindent {PACS number(s): 12.15.Ji, 12.15.Lk,
14.80.Bn }

\end{abstract}
\maketitle

\section{Introduction}

The recent discovery of a Standard Model(SM) Higgs-like boson
with mass near $125${GeV} by ATLAS \cite{Atlas} and CMS
\cite{Cms} Collaborations in Large Hadron Collider(LHC) experiment
stirs the world with a great interest in high energy physics.
Following, to identify whether it is the SM Higgs boson or not, and
to understand further its nature are of the upmost
goals in high energy physics. In order to ascertain the new finding
is just the SM Higgs boson, but not others, more measurements on its
decays are necessary. Therefore, to hunt for more LHC experiment
accessible Higgs processes is an important task for theoretical study.

In ATLAS and CMS experiments, the SM Higgs-like particle was
observed via its five decay channels, i.e., to $\gamma \gamma$,
$Z^{(*)} Z$, $W^{(*)} W$, $\tau^+ \tau^-$ and $b \bar{b}$. All these
decay processes are, at least intermediately, in two-body decay mode.
In this paper, we propose to study a complementary channel,
the three-body lepton-pair radiative decay processes $H\rightarrow
\ell\bar{\ell}\gamma$ ($\ell$ stands for leptons), to reconstruct
the SM Higgs-like neutral boson around $125${GeV}. Due to the
enhancement induced by internal heavy quarks and gauge bosons,
we find that these three-body decay processes will not be suppressed much
relative to the pure lepton-pair decay modes, and are observable with
the data collected in 2011 and 2012.

Before the LHC run, some of the radiative lepton-pair decay processes
were analyzed \cite{repko,ali,ana}, where \cite{repko} considered only 
the light fermions and hence neglected the tree diagram contribution; 
\cite{ali} evaluated the Higgs boson radiative decay to muon pair with 
Next-to-Leading Order(NLO) corrections; the calculation in \cite{ana} was incomplete; 
and in all these analyses the realistic LHC
experimental conditions were not fully considered. In this work,
we calculate completely the SM Higgs radiative decays to lepton pairs
under the helicity basis, which may issue more information about the
Higgs boson. Various physical cuts for reconstructing the Higgs
boson in LHC experiment are applied in the calculation. The partial
decay width with respect to the invariant mass of the muon pair is
given in order to disentangle the SM Higgs with Higgs bosons
in other models \cite{qiao}. The $H\rightarrow \tau^+
{\tau^-}\gamma$ process is also calculated beyond the Born level, where
the tau mass is non-negligible. Although in this process the leading order tree
diagrams contribute dominantly, we find the NLO contribution is
very helpful in experimental measurement due to the Z-pole effect, which
may greatly suppress the background.

Recently, Gastmans, Wu and Wu pointed out that the
dimensional regularization scheme may possess some shortcomings in
the calculation of Higgs to diphoton decays
\cite{ttwu1,ttwu2}. More explicitly, they believe that the W-loop
contribution should decouple when taking the zero-mass limit for W
boson. However, it is not the case in real calculation in dimensional
regularization scheme \cite{pave}. Their works stimulated many discussions
on this point in the literature
\cite{ktchao,CenZhang,YLWu,shifman,dedes}. Since in this work we
also encounter the triangle-loop diagrams appearing in the Higgs
to diphoton decays process, we will check again whether the W-boson
decouples or not in the massless limit.

The remainder of the paper is arranged as follows. In Sec. II, the
analytical formulae for the concerned processes are presented and
the W-boson decouple assumption is also confronted. In Sec. III, we
give out our numerical results and related phenomenological
discussions. The last section is devoted to conclusions.

\section{Formulae}

\subsection{The Born amplitude}
\begin{figure}[h]
\begin{center}
\includegraphics[scale=0.6]{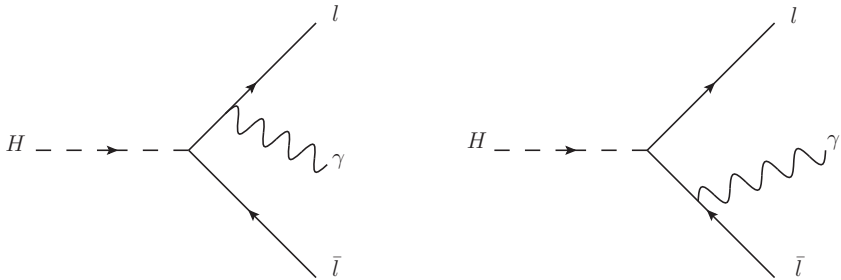}
\caption{The tree level diagrams of Higgs boson radiative decays to
lepton pairs.}
\end{center}
\label{born-fig}
\end{figure}

At tree level, there are only two Feynman Diagrams for the Higgs boson
radiative decays to lepton pairs, which are shown in Fig. 1. For the
convenience of following discussions, the momenta of particles are
reassigned as: $p_1=p_H$, $p_2=p_{\gamma}$, $k_1=p_{\ell}$, and
$k_2=p_{\bar{\ell}}$. The Mandelstam invariants are defined as
$s=(k_1+k_2)^2$, $t=(k_2+p_2)^2$,  $u=(k_1+p_2)^2$, and
$s+t+u=2m_\ell^2+m_H^2$. After doing some simplification, the tree
level amplitude can be expressed as
\begin{align} \mathcal {M}_{tree}\,&=\,\frac{e^2m_{l}}{2m_Ws_W}
\bigg\{\frac{\bar{u}(k_1)\not\!{p_2}\not\!{\varepsilon}
{v}(k_2)-2\varepsilon\cdot
k_1\bar{u}(k_1)(k_2)}{u-m_{\ell}^2}\nonumber\\
&+\frac{\bar{u}(k_1)\not\!{p_2}\not\!{\varepsilon}
{v}(k_2)+2\varepsilon\cdot
k_2\bar{u}(k_1){v}(k_2)}{t-m_{\ell}^2}\bigg\}\; .
\end{align}

\subsection{The One-Loop Amplitudes}

\begin{figure}[h]
\begin{center}
\includegraphics[scale=0.5]{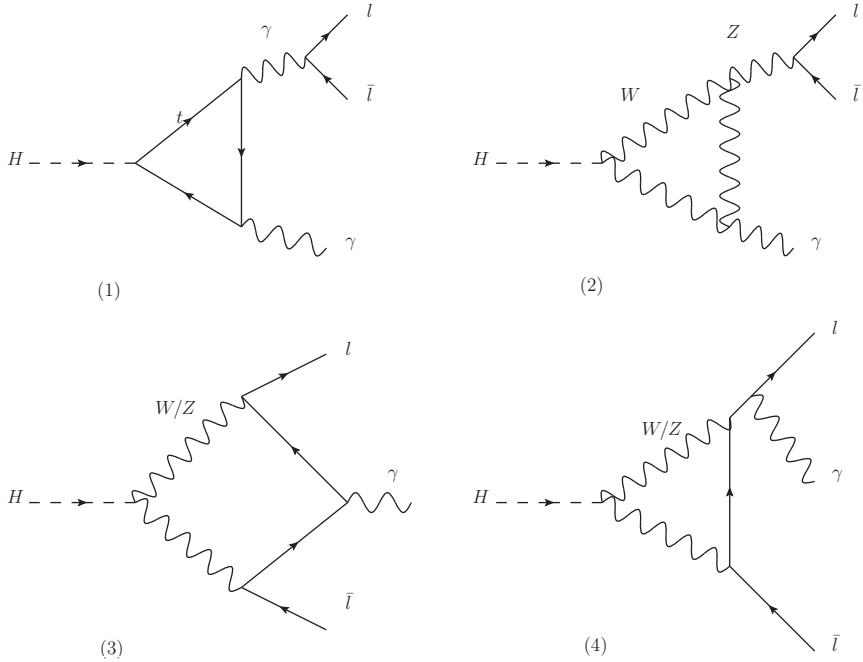}
\caption{The typical loop diagrams of Higgs boson radiative decays to
lepton pairs.}
\end{center}
\end{figure}

For the concerned processes, the next-to-leading order
contribution is very important for at least two reasons. First, the
large couplings of Higgs boson with heavy fermions and gauge bosons
greatly enhance the decay rates at NLO. Secondly, the intermediate
Z-boson decays may greatly reduce the backgrounds and enable
the experimental measurement on signals more transparent.

The typical one-loop Feynman diagrams are shown in Fig. 2. In our
calculation, both tree level and NLO amplitudes are generated by
virtue of the Mathematica package FeynArts \cite{Feynarts}. The
analytical calculation of the amplitudes proceeds with the help of
package FeynCalc \cite{Feyncalc}. Through out the calculation, the
Feynman-'t Hooft gauge is adopted, and for the convenience of calculation
and discussion, the one-loop Feynman diagrams
are classified into four groups, i.e., 1) the group of triangle
diagrams with virtual photon directly coupling to $\ell\bar{\ell}$;
2)the group of triangle diagrams with virtual Z-boson directly
coupling to $\ell\bar{\ell}$; 3) the group of W- or Z-box
diagrams, and 4) the group of triangle diagrams with a photon emitting
from leptons.

In Feynman-'t Hooft  gauge, there are 28 triangle diagrams in the
first group. After doing some algebraic reduction and simplification
the amplitude in this group can be expressed as
\bqa \mathcal {M}_{\gamma}=\mathcal {M}_{\gamma}^t+\mathcal
{M}_{\gamma}^W \; ,\eqa
where the amplitudes for top quark and W boson(including Goldstone
and ghost) induced diagrams read
\begin{align}
\mathcal {M}_{\gamma}^t\,&=\,\frac{-m_t^2e^4N_c}{18\pi^2m_W\sin
\theta_Ws(m_H^2-s)^2}\big\{(m_H^2-s)((m_H^2-s
-4m_t^2)\times\nonumber\\
&C_0(m_H^2,0,s,m_t^2,m_t^2,m_t^2)-2)+2s(B_0(s,m_t^2,m_t^2)
-B_0(m_H^2,m_t^2,m_t^2))\big\}\times\nonumber\\
&\big((m_H^2 -s)\bar{u}(k_1)\not\!{\varepsilon}
{v}(k_2)-2\varepsilon\cdot
p_1\bar{u}(k_1)\not\!{p_2}{v}(k_2)\big)\;,
\end{align}
and
\begin{align}
\mathcal {M}_{\gamma}^W\,&=\,\frac{e^4}{16\pi^2m_W\sin
\theta_Ws(m_H^2-s)^2}\big\{(m_H^2-s)(m_W^2(m_H^2-s)(6m_H^2
-12m_W^2-5s)\times\nonumber\\
&C_0(m_H^2,0,s,m_t^2,m_t^2,m_t^2)-(m_H^2+6m_W^2)
(sB_0(m_H^2,m_W^2,m_W^2)-sB_0(s,m_W^2,m_W^2)\nonumber\\
&+m_H^2-s))\bar{u}(k_1)\not\!{\varepsilon}
{v}(k_2)-2(-2m_W^2(m_H^2-s)(-3m_W^2+6m_H +4s)\times\nonumber\\
& C_0(m_H^2,0,s,m_t^2,m_t^2,m_t^2)-(m_H^2+6m_W^2)
(sB_0(m_H^2,m_W^2,m_W^2)-sB_0(s,m_W^2,m_W^2)\nonumber\\
&+m_H^2-s))\varepsilon\cdot p_1
\bar{u}(k_1)\not\!{p_2}{v}(k_2)\big\}\; ,
\end{align}
respectively. Since the bottom quark mass is about two orders
smaller than the top quark mass, and the electric charge of bottom
quark is half of top quark, it is reasonable to neglect the
contribution of b-quark loop diagram. Note that the above amplitudes
contain neither ultraviolet nor infrared divergences. The $C_0$ and $B_0$
functions are defined the same as in Ref. \cite{looptools}. The value of
the $C_0$ function and the finite part of the $B_0$ function can be readily
obtained by Looptools \cite{looptools}, and they have also been
checked with analytical results.

To see whether the W-loop decouples from the decay amplitude or not, we take
a cut on the virtual photon propagators in the group one diagrams, i.e., take the
zero limit for Mandelstam variable s, and then factorize out the
photon polarization vector. After doing these, the amplitude $\mathcal
{M}_{\gamma}^W$ turns to be
\begin{align}
\mathcal {M}_{s=0}^{cut}\,&=\,-\frac{i e^3}{16\pi^2m_W
\sin\theta_Wm_H^2}((k_1+k_2)^{\mu}p_2^{\nu}-g^{\mu\nu}m_H^2)\big\{
2m_H^2+12m_W^2+12m_W^2(2m_W^2\nonumber\\
&-m_H^2)C_0(m_H^2,0,0,m_W^2,m_W^2,m_W^2) \}\; ,
\end{align}
which is in agreement with results obtained in Refs. \cite{CenZhang,ktchao}.
This confirms that the W-boson loop contribution will not
decouple from the amplitude in the zero mass limit. Moreover, while taking the
zero limit for $s$ in top-loop amplitude, we get the same result as (4.11)
of Ref. \cite{ktchao}, which in some sense convinces us of our calculation.

Similar to $\mathcal{M}_{\gamma}$, there are also 28 triangle
diagrams in group two with virtual Z boson directly coupling to the
lepton pairs. Here, the ultraviolet divergence appears, which
can be renormalized by adding proper counter terms in the
on-shell renormalization scheme as given in Ref. \cite{denner}. In the
end, we get the finite amplitude as
\bqa \mathcal {M}_{Z}=\mathcal {M}_{Z}^t+\mathcal {M}_{Z}^W\; , \eqa
where the sub-amplitudes for top-quark and W-boson(including Goldstone
and ghost) induced loop diagrams read
\begin{align}
\mathcal {M}_{Z}^t\,&=\,\frac{m_t^2e^4N_c}{576\pi^2\sin^3
\theta_W\cos^2\theta_W
m_W(m_H^2-s)^2(s-m_Z^2+i\Gamma_Zm_Z)}(8\sin^2\theta_W-3)\big\{\nonumber\\
&(m_H^2-s)((s-m_H^2+4m_t^2)C_0(m_H^2,0,s,m_t^2,m_t^2,m_t^2)+2)
+2sB_0(m_H^2,m_t^2,m_t^2)\nonumber\\
&-2sB_0(s,m_t^2,m_t^2)\big\}\big((m_H^2
-s)(\bar{u}(k_1)\not\!{\varepsilon}\gamma^5
{v}(k_2)+(4\sin^2\theta_W-1)\bar{u}(k_1)\not\!{\varepsilon}
{v}(k_2))\nonumber\\
&-2\varepsilon\cdot p_1(\bar{u}(k_1)\not\!{p_2}\gamma^5
{v}(k_2)+(4\sin^2\theta_W-1)\bar{u}(k_1)\not\!{p_2}
{v}(k_2))\big)\nonumber\\
&-\frac{i m_t^2e^4N_c}{96\pi^2\sin^3\theta_W\cos^2\theta_W
m_W(m_H^2-s)(s-m_Z^2+i\Gamma_Zm_Z)}\big\{\nonumber\\
& \big(\epsilon^{\mu\nu\rho\sigma}p_{1\rho}
p_{2\sigma}((m_H^2-s)C_0(m_H^2,0,s,m_t^2,m_t^2,m_t^2)
+2B_0(m_H^2,m_t^2,m_t^2)\nonumber\\
&-2B_0(s,m_t^2,m_t^2))\big)
\big\}\varepsilon_{\mu}\big(\bar{u}(k_1)\gamma_{\nu}\gamma^5{v}(k_2)
+(4\sin^2\theta_W-1)\bar{u}(k_1)\gamma_{\nu}{v}(k_2)\big)\;,
\end{align}
\begin{eqnarray}
\mathcal {M}_{Z}^W&&=\frac{e^4}{128\pi^2m_W\sin^3\theta_W\cos^2
\theta_W(m_H^2-s)(s-m_Z^2+i\Gamma_zm_Z)}\big\{\nonumber\\
&&2(m_H^2-s)((3s-5m_H^2+13m_W^2)\cos^2\theta_W+(m_H^2
-2s+m_W^2)\sin^2\theta_W)\times\nonumber\\
&&m_W^2C_0(m_H^2,0,s,m_W^2,m_W^2,m_W^2)
+(\cos^2\theta_W(m_H^2+10m_W^2)\nonumber\\
&&-(m_H^2+2m_W^2)\sin^2\theta_W)(m_H^2-s
+sB_0(m_H^2,m_W^2,m_W^2)\nonumber\\
&&-sB_0(s,m_W^2,m_W^2))\big\}(\bar{u}(k_1)\not\!{\varepsilon}\gamma^5
{v}(k_2)+(4\sin^2\theta_W-1)\bar{u}(k_1)\not\!{\varepsilon}
{v}(k_2))\nonumber\\
&&+\frac{e^4}{64\pi^2m_W\sin^3\theta_W\cos^2
\theta_W(m_H^2-s)^2(s-m_Z^2+i\Gamma_zm_Z)}\big\{\nonumber\\
&&-2(m_H^2-s)((6s-5m_H^2+10m_W^2)\cos^2\theta_W+(m_H^2
-2s-2m_W^2)\sin^2\theta_W)\times\nonumber\\
&&m_W^2C_0(m_H^2,0,s,m_W^2,m_W^2,m_W^2)
-(\cos^2\theta_W(m_H^2+10m_W^2)\nonumber\\
&&-(m_H^2+2m_W^2)\sin^2\theta_W)(m_H^2-s
+sB_0(m_H^2,m_W^2,m_W^2)\nonumber\\
&&-sB_0(s,m_W^2,m_W^2))\big\}\varepsilon\cdot
p_1(\bar{u}(k_1)\not\!{p_2}\gamma^5
{v}(k_2)+(4\sin^2\theta_W-1)\bar{u}(k_1)\not\!{p_2} {v}(k_2))\nonumber\\
&&+\frac{m_\ell m_We^4}{64\pi^2\sin^3\theta_W\cos^2
\theta_W(m_H^2-s)(s-m_Z^2+i\Gamma_zm_Z)}
\big\{\nonumber\\
&&2(m_H^2-s)(2\cos^2\theta_W-\sin^2\theta_W)
C_0(m_H^2,0,s,m_W^2,m_W^2,m_W^2)\nonumber\\
&& +B_0(s,m_W^2,m_W^2)-B_0(m_H^2,m_W^2,m_W^2)\big\}\varepsilon\cdot
p_1 \bar{u}(k_1)\gamma^5 {v}(k_2)\; ,
\end{eqnarray}
respectively. In the limit of Z being on mass shell, we obtain the known
decay amplitude for $H\rightarrow Z\gamma$, and our analytic
result agrees with Ref. \cite{repko}.

The calculation procedure for box diagrams is similar to what for the
triangle diagrams, except it becomes more tedious and complicate. However,
the analytic result for box diagrams is too lengthy to be presented in the paper.

Note that since the direct coupling of Higgs or Goldstone bosons to leptons
are suppressed by factor of $\frac{m_\ell}{m_H}$, which is
much less than one for light leptons, contributions of the corresponding diagrams
to the final result are negligibly small.
However, for $\tau$ lepton, those terms are kept due to its relatively large mass.

\subsection{Helicity amplitudes}

Notice that the helicity information on final states may tell more on
parent particles, we also perform the calculation on the helicity
basis for final states. Employing the helicity method given in Refs.
\cite{helicity1,helicity2}, by introducing a light-like momentum
$q_0$ and a space-like vector $q_1$ with constraints $q_0\cdot
q_1=0$ and $q_1\cdot q_1=-1$, the helicity amplitudes can be then
constructed as
\begin{eqnarray}
&&\mathcal {M}_{ss^\prime\lambda}= N_0
\mathrm{Tr}[(\not\!{k_2}-m_\ell)
(1-\gamma^5)\not\!{q_0}(\not\!{k_1}+ m_\ell)
\mathcal {A}_\lambda]\; ,\nonumber \\
&& \mathcal {M}_{-s-s^\prime\lambda}= N_0
\mathrm{Tr}[(\not\!{k_2}-m_\ell)
(1+\gamma^5)\not\!{q_0}(\not\!{k_1}+ m_\ell)
\mathcal {A}_\lambda]\; , \nonumber \\
&& \mathcal {M}_{-ss^\prime\lambda} = N_0\mathrm{
Tr}[(\not\!{k_2}-m_\ell) (1-\gamma^5)\not\!{q_0}\not\!{q_1}
(\not\!{k_1}+m_\ell)\mathcal {A}_\lambda]\; ,\nonumber \\
&&\mathcal {M}_{s-s^\prime\lambda}= N_0
\mathrm{Tr}[(\not\!{k_2}-m_\ell)
\not\!{q_1}(1-\gamma^5)\not\!{q_0}(\not\!{k_1}+m_\ell)\mathcal
{A}_\lambda]\; .
\end{eqnarray}
Here, $N_0=1/\sqrt{16(q_0\cdot k_1)(q_0\cdot k_2)}$; $\mathcal
{A}_\lambda=\mathcal {A}^{tree}_\lambda+\mathcal
{A}^{loop}_\lambda$, which can be readily obtained after removing the
spinors out from the above mentioned amplitude; $s=1/2$ and
$s^\prime=1/2$ denote for the spin projections of $\ell$ and
$\bar{\ell}$ respectively; $\lambda=\pm1$ represents the polarizations
of the photon. Hence, in the end there will be eight different kinds of
helicity amplitudes.

In the helicity method, the total unpolarized matrix element squared is
just the sum of individual polarized matrix element squared, i.e.,
\begin{equation}
\label{sum}
    |\mathcal{M}|^2=|\mathcal{M}_{ss^\prime\lambda}|^2+
    |\mathcal{M}_{-s-s^\prime\lambda}|^2+|
    \mathcal{M}_{-ss^\prime\lambda}|^2+|\mathcal
    {M}_{s-s^\prime\lambda}|^2\; .
\end{equation}
In practice of our calculation, we find that the helicity method gives the
same result of the total unpolarized matrix element squared as the traditional one,
but the former provides more information about the physical process
and in some case brings convenience in the calculation.

\section{the numerical calculation}

\subsection{Input parameters and physical conditions}
In numerical evaluation, the relevant inputs are taken as \cite{PDG,
Higgs-cs}
\begin{align}
&m_t=172.0 ~\mathrm{GeV},\; m_W=80.39 ~\mathrm{GeV},\; \alpha(m_Z)=1/128,\;
\Gamma_Z=2.48 ~\mathrm{GeV}, \nonumber\\
&m_Z=91.18 ~\mathrm{GeV},\; m_{\mu}=0.105 ~\mathrm{GeV},\;
m_e=0.51 ~\mathrm{MeV},\; m_\tau=1.776 ~\mathrm{GeV}\; .
\label{inputs}
\end{align}
The SM Higgs mass is taken to be $125.5$ {GeV} in our numerical evaluation,
which is around the mass of the Higgs-like boson recently observed by ATLAS
and CMS Collaborations \cite{Atlas,Cms}.

Formally, the three-body partial decay width can be expressed as
\bqa d\Gamma=\frac{1}{(2\pi)^3}\frac{1}{32m_H^3}|\mathcal {M}|^2\ ds\
dt\; .\eqa
Before imposing any physical cuts, the Mandelstam variables $s$ and $t$,
defined previously, vary in space of
\bqa s_{max} = m_H^2\; ,\ s_{min}=4m_\ell^2\; , \eqa
and
\bqa t_{max}=\frac{m_H^4}{4s}-(\sqrt{\frac{s}{4}-m_\ell^2}-
\frac{m_H^2-s}{2\sqrt{s}})^2\; ,\
t_{min}=\frac{m_H^4}{4s}-(\sqrt{\frac{s}{4}-m_\ell^2}
+\frac{m_H^2-s}{2\sqrt{s}})^2\; , \eqa
respectively.

Considering of the experimental constraints, certain physical cuts should
be imposed on the phase space. In our calculation three sets of physical cuts
are taken for $s$, $t$, $u$ and momenta of final states, as presented in Table
\ref{tab:cuts}. That is $s\geq
(m^2_{\ell\bar{\ell}})_{\mathrm{cut}}$, $t\geq
(m^2_{\bar{\ell}\gamma})_{\mathrm{cut}}$, $u\geq
(m^2_{\ell\gamma})_{\mathrm{cut}}$, $E_\ell \geq
(E_\ell)_{\mathrm{cut}}$, $E_{\bar{\ell}} \geq
(E_{\bar{\ell}})_{\mathrm{cut}}$, and $E_\gamma \geq
(E_\gamma)_{\mathrm{cut}}$, where the last three constraints are
imposed in the center of mass system of Higgs boson. These cuts, especially
the cut III, facilitates the experiment measurement on photon and
leptons. Note that for Higgs radiative decay to $\tau$ pair, the low limits
of lepton energies in cut I are taken to be the $\tau$ mass,
but rather the values given in Table I.
In addition of the cuts in the table, in practice
the experimental conditions are also taken into account \cite{cms1}. That is,
for CMS experiment the transverse momenta of photon and leptons should be
larger than 15 {GeV} and 10 {GeV}(or 20 {GeV}) respectively, and the pseudo
rapidity in experiment measurement is constrained within the
fiducial volume of $|\eta|<2.5$.

\begin{table}[thb]
\caption{\label{tab:cuts} Selection cuts  in numerical evaluation.}
\begin{center}
\begin{tabular}{|c|c|c|c|c|c|c|}
\hline
  & $(m^2_{\ell\bar{\ell}})_{\mathrm{cut}}$  &
  $(m^2_{\bar{\ell}\gamma})_{\mathrm{cut}}$  &
 $(m^2_{\ell\gamma})_{\mathrm{cut}}$  &
 $(E_\ell)_{\mathrm{cut}}~(\mathrm{GeV})$ &
 $(E_{\bar{\ell}})_{\mathrm{cut}}~(\mathrm{GeV})$
  & $(E_\gamma)_{\mathrm{cut}}~(\mathrm{GeV})$ \\
\hline cut I & $25m_\mu^2$ &$ 25m_\mu^2$   &$25m_\mu^2$   & 1
 &  1&  1  \\
\hline
cut II  & $50m_\mu^2$ &$ 50m_\mu^2$   &$50m_\mu^2$ & 6& 6  &6\\
\hline cut III  &$75m_\mu^2$ &$75m_\mu^2$   &$75m_\mu^2$  &
10 & 10 & 10 \\
\hline
\end{tabular}
\end{center}
\end{table}

\subsection{Numerical results and discussions}

With the analytic expressions obtained, the input parameters given
and the physical conditions selected in preceding sections, the decay
widths and branching fractions of a $125.5$ {GeV} SM Higgs boson
radiative decays to lepton pairs can be readily obtained, which are
shown in Table \ref{tab:results1} and Table \ref{tab:results2}.
For these processes, the backgrounds are huge, of which the dominant ones come
from the Drell-Yan + ISR(initial state radiation)
and Drell-Yan + FSR(final state radiation) processes \cite{cms1,keung}.
The backgrounds are evaluated with the help of the package CalcHep
\cite{calchep}, and the results are presented in Table IV.
For illustration, the invariant mass distributions
of the decay mode $H \rightarrow \mu^+\mu^-\gamma$ are shown in
Figs. 3 and 4, for summed and individual helicity cases,
respectively. For Higgs to $e^+e^-\gamma$ decay mode,
result shows that contributions of tree and tree-loop interference terms are negligible,
as mentioned in Ref. \cite{ali}. Hence, the
invariant mass distribution of $H \rightarrow e^+ e^-\gamma$ process
tends to be as the NLO mass distribution of
$H\rightarrow\mu^+\mu^-\gamma$ decay mode, that is the subtraction
of the total contribution by the tree level contribution as shown in
Fig. 3. For $H\rightarrow \tau^+ \tau^-\gamma$ process, the dominant
contribution comes from the leading order diagrams, due to the
relatively large Higgs-$\tau$ coupling. However, the subleading
contribution for $H\rightarrow \tau^+ \tau^-\gamma$ process from the
internal Z conversion to $\tau$ pair tends to be meaningful with the
help of Z-pole veto in experimental measurement.

\begin{table}[h]
\caption{\label{tab:results1} The decay widths and branching
fractions with various Cuts. Here, the mass of Higgs boson is taken
to be $125.5${GeV}.}

\begin{center}
\vspace{-0.7cm}
\begin{tabular}{|c|c|c|c|}
\hline
  &  $ \Gamma_{e^+e^-\gamma}(10^{-7}\mathrm{GeV})/Br(10^{-4})$ &
  $\Gamma_{\mu^+\mu^-\gamma}(10^{-7}\mathrm{GeV})/Br(10^{-4}) $
  &  $\Gamma_{\tau^+\tau^-\gamma}(10^{-6}\mathrm{GeV})/Br(10^{-3}) $ \\
\hline cut I &  4.29/1.04
& 5.67/1.38 &  33.4/8.10  \\
\hline
cut II  &3.89/0.94 & 4.57/1.11  & 16.1/3.90 \\
\hline cut III  &
3.61/0.88 & 4.09/0.99 & 11.5/2.79 \\
\hline
\end{tabular}
\end{center}
\end{table}
\begin{table}[thb]
\caption{\label{tab:results2} The decay widths and branching
fractions with different transverse momentum cuts for leptons.
The minimum transverse momentum for photon is set to be 15 {GeV},
the pseudo rapidity $|\eta|$ is constrained within $2.5$,
and the invariant mass cuts are taken to be the Cut III of Table I.}
\begin{center}
\begin{tabular}{|c|c|c|c|}
\hline
&  $ \Gamma_{e^+e^-\gamma}(10^{-7}\mathrm{GeV})/Br(10^{-4})$
& $\Gamma_{\mu^+\mu^-\gamma}(10^{-7}\mathrm{GeV})/Br(10^{-4}) $ \\
\hline $p^\ell_{\perp,\mathrm{cut}}=10${GeV} & 2.73/0.66
&  2.94/0.71   \\
\hline
$p^\ell_{\perp,\mathrm{cut}}=20${GeV} &  2.11/0.51 &  2.28/0.55   \\
\hline
\end{tabular}
\end{center}
\end{table}
\begin{table}[thb]
\caption{\label{tab:background} The background of Drell-Yan plus ISR
or FSR photon. The cuts are of the same as in signal production in Table III. The
invariant mass of $\ell\bar{\ell}\gamma$ lies in the scope of
$115-180$ {GeV} as in Ref. \cite{cms1}.}
\begin{center}
\begin{tabular}{|c|c|c|c|c|c|}
\hline & ISR $p^\ell_{\perp,\mathrm{cut}}=10${GeV}
& ISR $p^\ell_{\perp,\mathrm{cut}}=20${GeV}  &
FSR $p^\ell_{\perp,\mathrm{cut}}=10${GeV} & FSR $p^\ell_{\perp,\mathrm{cut}}=20${GeV} \\
\hline $\sigma(e^+e^-\gamma)$ & 3.24$\times10^{-1}$pb
&  2.78$\times10^{-1}$pb & 1.27$\times10^{-1}$pb & 1.02$\times10^{-1}$pb \\
\hline
$\sigma(\mu^+\mu^-\gamma)$ & 3.26$\times10^{-1}$pb
 & 2.8$\times10^{-1}$pb & 1.28$\times10^{-1}$pb & 1.02$\times10^{-1}$pb \\
\hline
\end{tabular}
\end{center}
\end{table}
Numerical evaluation indicates that the contributions of Box
and triangle diagrams, in which the photon emits from one of the
fermions, are tiny in comparison with what from the first two
groups. Furthermore, among the loop diagrams those with a W triangle
are primary for the decay processes. Since the triangle diagrams
in our calculation have similar property to what of the Higgs to
two photon process, we also check the W decoupling assumption by
taking the zero mass limit, and find that W-loop contribution will
not decouple in the dimensional regularization scheme.

Results show that the sum of the branching ratios of
$H\rightarrow\mu^+\mu^-\gamma$ and $H\rightarrow e^+e^-\gamma$
processes with cut I is about two times bigger than that of
$H\rightarrow \ell\bar{\ell}\ell\bar{\ell}(\ell=\mu~or~e)$ processes
\cite{Higgs-cs}, where no cut is employed, and also lager than the
branching ratio of $H\rightarrow\mu^+\mu^-$
decay channel. Since $H\rightarrow
\ell\bar{\ell}\ell\bar{\ell}(\ell=\mu~or~e)$ processes have already
been observed in LHC experiment, the Higgs radiative decays to
lepton pairs are very likely to be measured.

\begin{table}[thb]
\caption{\label{tab:helicity results} The helicity dependent decay
widths and branching fractions with Cut I.}
\begin{center}
\vspace{-0.7cm}
\begin{tabular}{|c|c|c|c|}
\hline
  &  $ \Gamma_{e^+e^-\gamma}(10^{-7}\mathrm{GeV})/Br(10^{-4})$ &
  $\Gamma_{\mu^+\mu^-\gamma}(10^{-7}\mathrm{GeV})/Br(10^{-4}) $
  &  $\Gamma_{\tau^+\tau^-\gamma}(10^{-6}\mathrm{GeV})/Br(10^{-3}) $ \\
\hline  flip &  4.29/1.04
& 4.30/1.04 &  0.3/0.07  \\
\hline
non-flip  & 0.00/0.00 & 1.37/0.34  &  13.1/3.19 \\
\hline
\end{tabular}
\end{center}
\end{table}

\begin{figure}[thb]
\begin{center}
\includegraphics[scale=0.5]{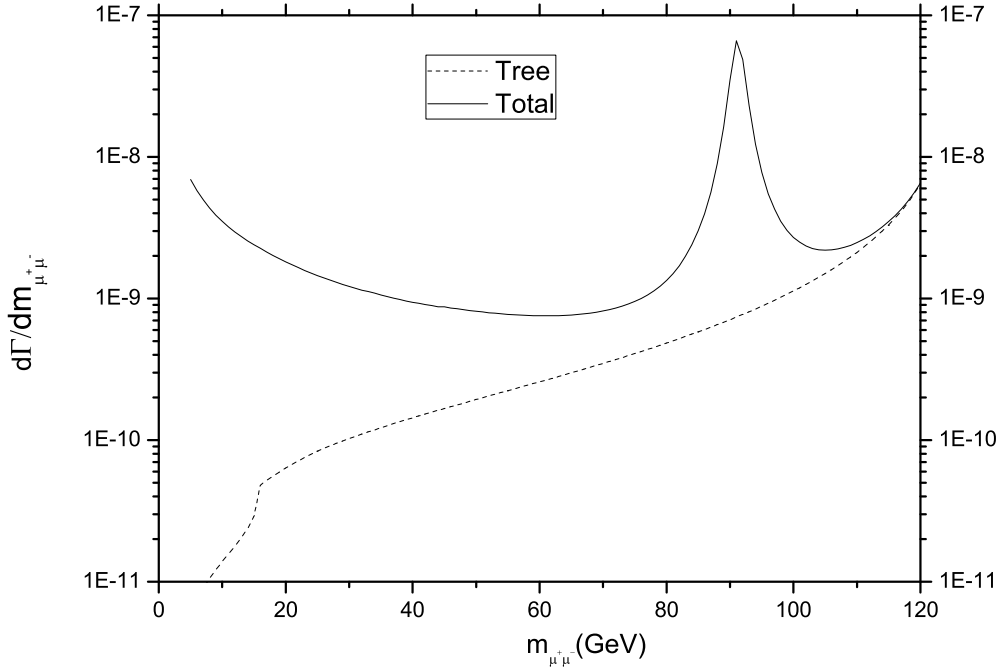}
\caption{ The $\mu\bar{\mu}$ invariant mass distributions of Higgs
decay to $\mu\bar{\mu}\gamma$. Here, the premise of Cut I is taken
and the Higgs mass is set to be $125.5$ {GeV}.}
\end{center}
\end{figure}

\begin{figure}[thb]
\begin{center}
\includegraphics[scale=0.5]{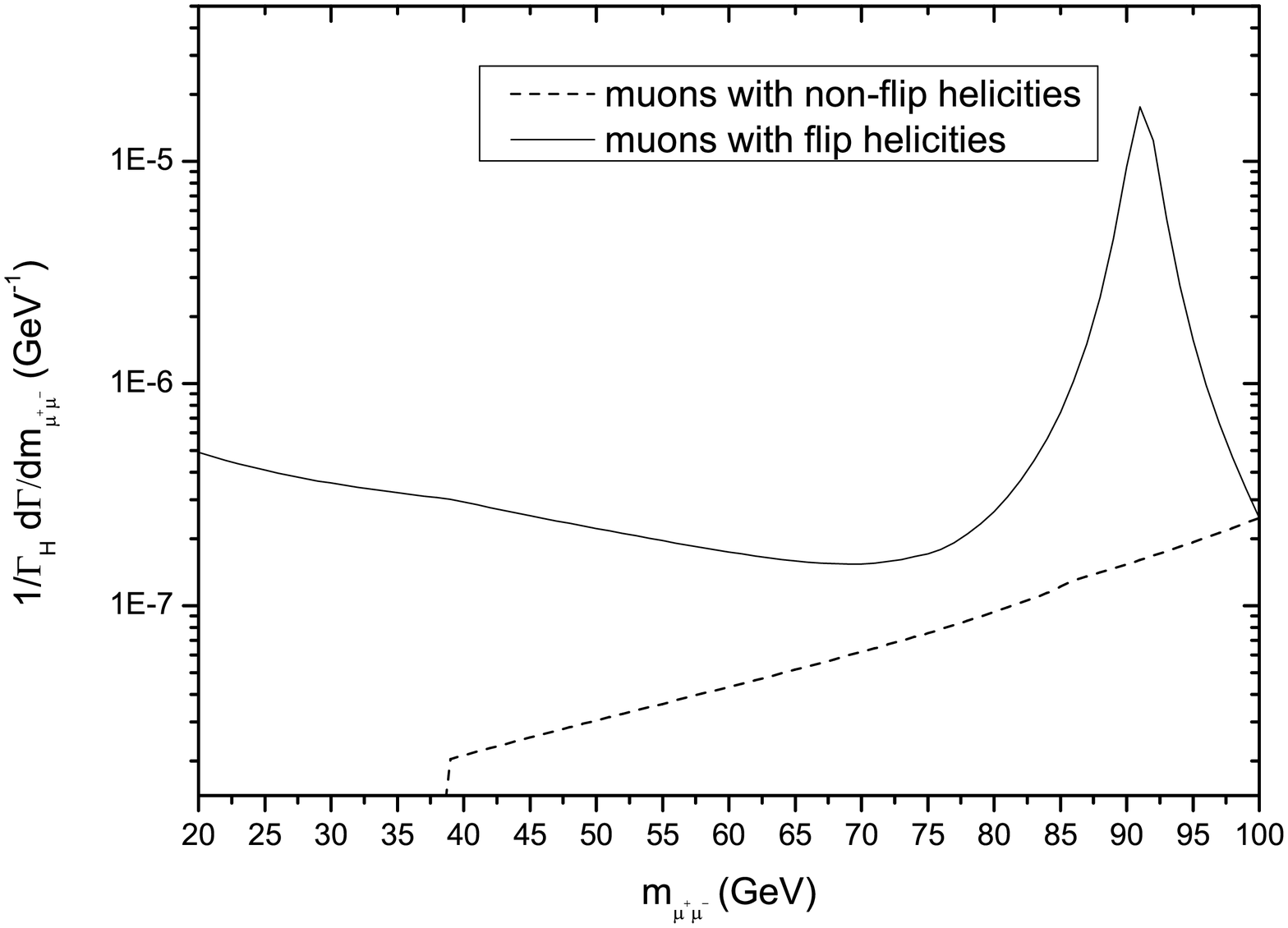}
\caption{ The $\mu\bar{\mu}$ invariant mass distributions of the Higgs
radiative decay to muon pair with different helicities,
the flipped (--+ and +--) and non-flipped(--~--and ++) cases. Here,
the premise of physical condition Cut II is adopted.}
\end{center}
\end{figure}

From Fig. 4 we notice that the contributions from muon and electron
helicity non-flip processes are negligible in Higgs radiative
decays. This is not a surprise, since in fact the helicity flip events
are mainly induced by the loop diagrams. For tree diagrams, however,
the largest contribution comes from the soft photon radiation where
muon pairs are produced back-to-back and photon is collinear to one
of them, which leads to non-flip helicities of the muon pairs. The
detailed helicity dependent results for decay width and branching
ratio are given in Table \ref{tab:helicity results}.

The backgrounds of Drell-Yan+ISR and
Drell-Yan+FSR are shown in Table IV. Among
Drell-Yan + ISR processes, $pp\to Z\gamma\to
\ell\bar{\ell}\gamma$ has a peak around the Z boson
mass shell. While for Drell-Yan + FSR processes, the value of
$m_{\ell\bar{\ell}\gamma}$ in scope of 115-180 {GeV} ensures the
intermediate particle to be virtual. Hence, for our concerned issue
the ISR background is bigger than that of FSR. From Table IV we notice that
these two main backgrounds are in order of $10^{-1}$pb with actual
experimental cuts for signal production. Furthermore, to enhance the
signal tagging efficiency and suppress the background, one can perform
the measurement on muon pair with flip helicities and invariant mass in
scope of $75$ {GeV} to $100$ {GeV}, where half of the Cut I events
with photon energy lying in $20$ {GeV} to $40$ {GeV} will
be produced because of the Z-pole effect as shown in Fig. 4.
Apart from the Drell-Yan background, there are still other processes,
like $gg\rightarrow \ell\bar{\ell}\gamma$, contribute to the background.
However, they are not evaluated in this work, due to not only detailed
background analyses need more works, but superficially these processes
are higher order ones.

\section{Conclusions\label{sec-con}}

In this paper, the Higgs radiative decays to lepton pairs processes
$H\rightarrow \ell\bar{\ell}\gamma
(\ell=e,\mu,\tau)$ are investigated up to next-to-leading order in
perturbative expansion in electroweak interaction. The branching
ratios and decay widths are calculated in polarization summed and separated
cases with certain physical constraints.
The physical Cuts are imposed in the calculation in order to
mimic the experimental conditions and promote the significance of
the signal. Numerical result shows that the
branching ratios of $H\rightarrow \ell\bar{\ell}\gamma (\ell=e,\mu)$
processes are at order of $10^{-4}$, some two times larger than that of
$H\rightarrow \gamma Z \rightarrow
\gamma\ell^+\ell^-(\ell=\mu~or~e)$ process, while the branching fraction
of $H\rightarrow \tau\bar{\tau}\gamma$ process is an order of magnitude higher.
Our calculation indicates that although the Z-pole effect is
overwhelming, the photon conversion to lepton pairs process is not
negligible, especially when the lepton pair with a minimum opening
angle. Since the Higgs radiative decays to electron and muon pairs
have branching ratios of order $10^{-4}$, by virtue of the helicity
measurement and Z-pole constraint, it is expected that the signals of
these processes will be observed in accumulated data or in the next
run of the LHC experiment, provided the Standard Model Higgs is
indeed about $125$ GeV.

Recently, the CMS Collaboration made a progress in measuring
$H\rightarrow \tau\bar{\tau}$ process \cite{Roger}.
Due to the large coupling of tau to Higgs boson,
unlike the electron or muon pair radiative production, the tau pair
radiative production in Higgs decay is mainly induced by the leading
order process in electroweak interaction.
However, the helicity measurement and Z-pole
constraint may also be helpful to measure the Higgs radiative decay
to tau pairs in NLO process.

Finally, we have also checked the W decoupling assumption in this
work by taking the zero mass limit of the W boson, and find that the
W-loop contribution does not decouple from the processes of Higgs radiative
decays to lepton pairs in the dimensional regularization scheme.

Note added: while this paper was submitted, Dicus and Repko post
a preprint on the reanalysis of Higgs radiative decay to electron-position
pair \cite{Dicus}.

\vskip 0.7cm
{\bf Acknowledgements}:
This work was supported in part by the
National Natural Science Foundation of China(NSFC) under the grants
10935012, 10821063 and 11175249.


\newpage
\begin{thebibliography}{99}

\bibitem{Atlas}
G. Aad, {\it et al.}, the ATLAS Collaboration,
 Phys.Lett. {\bf B 716},  1(2012).
\bibitem{Cms}
S. Chatrchyan, {\it et al.}, the CMS Collaboration, Phys.Lett. {\bf
B716}, 30(2012).
\bibitem{repko} A. Aabbasabadi, D. B. Chao, D. A. Dicus, and W.
W. Repko, Phys. Rev. {\bf D55}, 5647(1997).
\bibitem{ali} A. Aabbasabadi, Wayne.
W. Repko, Phys. Rev. {\bf D62}, 054025 (2000).
\bibitem{ana} Ana Firan and Ryszard Stroynowski,
Phys.\ Rev.\ {\bf D76}, 057301 (2007).
\bibitem{qiao} Chong-Sheng Li, Shou-Hua Zhu, and Cong-Feng Qiao,
Phys. Rev. {\bf D57}, 6928 (1998).
\bibitem{ttwu1} R. Gastmans, S. L. Wu, T. T.
Wu, [arXiv:hep-ph/1108.5322].
\bibitem{ttwu2} R. Gastmans, S. L. Wu, T. T.
Wu, [arXiv:hep-ph/1108.5872].
\bibitem{pave} G. Passarino and M. J. G. Veltman, Nucl. Phys.
{\bf B160}, 151 (1979).
\bibitem{CenZhang} William J. Marciano, Cen Zhang,
Scott Willenbrock, Phys. Rev. {\bf D85}, 013002 (2012)
\bibitem{ktchao} Hua-Sheng Shao, Yu-Jie Zhang, Kuang-Ta Chao, JHEP
01, 053(2012)
\bibitem{YLWu} Da Huang, Yong Tang, Yue-Liang Wu, Commun. Theor. Phys.
57, 427(2012)¡£
\bibitem{shifman} M. Shifman, A. Vainshtein, M.B. Voloshin, V.
Zakharov. Phys. Rev. {\bf D85}, 013015 (2012)
\bibitem{dedes} Athanasios Dedes, Kristaq Suxho,
[arXiv:hep-ph/1210.0141].
\bibitem{Feynarts} T. Hahn, Comput. Phys. Commun. {\bf
140}, 418 (2001).
\bibitem{Feyncalc} R. Mertig, M. Bohm and
A. Denner, Comp. Phys. Comm. \textbf{64}, 345 (1991).
\bibitem{looptools} T. Hahn and M. P$\acute{\text{e}}$rez-Victoria,
Comput. Phys. Commun. {\bf 118}, 153 (1999).
\bibitem{denner} A. Denner, Fortschr. Phys. {\bf 41}, 307(1993).
\bibitem{helicity1} R. Kleiss and W. J. Stirling, Nucl. Phys. {\bf B262},
245(1985); Zhan Xu, Da-Hua Zhang and Lee Chang, Nucl.Phys. {\bf
B291}, 392 (1987).
\bibitem{helicity2} Cong-Feng Qiao, Phys. Rev. {\bf D67},
097503 (2003); C.-H. Chang, J.-X. Wang and X.-G. Wu, Phys.Rev. {\bf
D77}, 014022 (2008).
\bibitem{PDG}
 K. Nakamura, {\it et al.}, Particle Data Group,
 J.\ Phys.\ G{\bf 37}, 075021 (2010).
 \bibitem{Higgs-cs}
S. Dittmaier, {\it et al.}, arXiv:1101.0593 [hep-ph].
\bibitem{cms1}  The CMS Collaboration,CMS-PAS-HIG-12-049.
\bibitem{keung}
J. S.Gainer, W.-Y. Keung, I. Low, and P. Schwaller, Phys. Rev. {\bf
D86}, 033010(2012).
\bibitem{calchep}
A. Pukhov, CalcHEP 2.3: MSSM, structure functions, event generation,
batchs, and generation of matrix elements for other packages (2004).
[arXiv:hep-ph/0412191].

\bibitem{Roger}
Roger Wolf, Search for the SM Higgs Boson in Di-$\tau$ Final States
at CMS, Hadron Collider Physics Symposium, Kyoto, 2012.

\bibitem{Dicus}
Duane A. Dicus and Wayne W. Repko, Phys.Rev. {\bf
D87}, 077301(2013). 

\end{thebibliography}
\end{document}